\documentclass[a4paper,prd,preprintnumbers,twocolumn,superscriptaddress,nofootinbib,amsmath,amssymb]{revtex4-2}

\usepackage[dvips]{graphics}
\usepackage{color,hyperref}
\usepackage{times}
\usepackage{mathrsfs}
\hypersetup{colorlinks=true,linkcolor=blue,citecolor=red,filecolor=magenta,urlcolor=blue}

\usepackage{enumitem}
\usepackage{graphicx,subfigure}
\usepackage{dcolumn}
\usepackage{bm}
\usepackage{color}

\begin{document}
\title{Long-lived quasinormal modes and  asymptotic tails of regular Schwarzschild-like black holes in the presence of a magnetic field}

    \author{Akbar Davlataliev}
    \email{akbar@astrin.uz}
    \affiliation{Inha University in Tashkent, Ziyolilar 9, Tashkent 100170, Uzbekistan}

	\author{Bakhtiyor~Narzilloev}
	\email{baxtiyor@astrin.uz}
  \affiliation{New Uzbekistan University, Movarounnahr str. 1, Tashkent 100000, Uzbekistan}
 \affiliation{Ulugh Beg Astronomical Institute, Astronomy St.  33, Tashkent 100052, Uzbekistan}
\affiliation{University of Tashkent for Applied Sciences, Str. Gavhar 1, Tashkent 100149, Uzbekistan}
     
	\author{Ibrar Hussain}
	\email{ibrar.hussain@seecs.nust.edu.pk}	
	\affiliation{School of Electrical Engineering and Computer Science, National University of Sciences and Technology, H-12, Islamabad, Pakistan}
	\author{Ahmadjon~Abdujabbarov}
	\email{ahmadjon@astrin.uz}
	\affiliation{Ulugh Beg Astronomical Institute, Astronomy St.  33, Tashkent 100052, Uzbekistan}
	\author{Bobomurat Ahmedov}
	\email{ahmedov@astrin.uz}
	\affiliation{New Uzbekistan University, Movarounnahr str. 1, Tashkent 100000, Uzbekistan}
	\affiliation{Institute of Fundamental and Applied Research, National Research University TIIAME, Kori Niyoziy 39, Tashkent 100000, Uzbekistan}

\date{\today}
\begin{abstract}
We analyze the evolution of perturbations of a (charged) massive scalar field near a regular Simpson-Visser black hole, allowing for a non-zero external magnetic field. We show that the damping rate of the quasinormal frequencies is strongly suppressed by both the magnetic field and the mass term, with indications that arbitrarily long-lived modes, or quasi-resonances, may exist in the spectrum. In the time domain, the quasinormal ringing transitions into slowly decaying oscillatory tails, which are qualitatively distinct from the massive tails observed in the absence of a magnetic field. For nonzero multipole and azimuthal numbers, the power-law envelope characteristic of cases without a magnetic field transforms into an oscillatory envelope that cannot be easily fitted with a simple analytical formula.
\end{abstract}

\maketitle

\section{Introduction}

Quasinormal modes (QNMs) \cite{Kokkotas:1999bd, Nollert:1999ji, Berti2009, Konoplya:2011qq} are fundamental features of black holes, representing the oscillatory decay of perturbations in their surrounding spacetime. These modes emerge from disturbances such as binary mergers, producing gravitational waves with characteristic frequencies that encode critical information about the mass and spin of black holes. Observations from LIGO, starting with GW150914, have confirmed the presence of these QNMs in gravitational wave signals, offering strong evidence to support the predictions of General Relativity (GR) \cite{Abbott2016, LIGOScientific:2020zkf, LIGOScientific:2017vwq, Isi2019}. Future observations with advanced detectors like LISA aim to refine our understanding, testing the no-hair theorem, and investigating deviations that could indicate new physics beyond GR \cite{Berti2009, Cardoso2021}.

Perturbations and the quasinormal spectrum of a massive or effectively massive field differ qualitatively from those of a massless field. First, the mass term significantly suppresses the damping rate of frequencies \cite{Simone:1991wn,Burikham:2017gdm,Wu:2015fwa,Dubinsky:2024hmn}, which can sometimes lead to the existence of arbitrarily long-lived modes, known as quasi-resonances \cite{Ohashi:2004wr,Konoplya:2004wg}. 

This phenomenon is quite broad and encompasses various spherically and axially symmetric black holes in four and higher dimensions, as well as fields with different spin \cite{Konoplya:2006br,Zhidenko:2006rs,Zhang:2018jgj,Konoplya:2017tvu,Bolokhov:2023bwm}. However, there are configurations in which quasi-resonances do not occur, despite considerable suppression of the decay rate \cite{Konoplya:2013rxa,Aragon:2020teq,Zinhailo:2024jzt}. Therefore, determining the existence of arbitrarily long-lived modes must be done on a case-by-case basis.

It is also noteworthy that initially massless fields may acquire an effective mass term due to various factors, such as tidal forces from extra dimensions in brane-world models \cite{PhysRevLett.94.121302,Boehmer:2009bmu} or the presence of an external magnetic field \cite{Konoplya:2007yy,Wu:2015fwa}, which may induce a strong superradiant instability \cite{Konoplya:2008hj,Brito:2014nja}. The influence of a magnetic field will be relevant in our analysis as well.

In the time domain, massive fields also exhibit distinctive decay characteristics. Although massless fields typically transition from quasinormal ringing to power-law decay, massive fields decay in a slow, oscillatory manner with a power-law envelope \cite{Hod:1998ra,Koyama:2001qw,Koyama:2001ee,Rogatko:2007zz,Moderski:2001tk,Gibbons:2008gg,Jing:2004zb,Konoplya:2006gq}.

These slowly decaying oscillatory tails were employed in \cite{Konoplya:2023fmh} to suggest that a massive graviton or other massive particles, whether in the framework of massive gravity or as an effective force, could contribute to the very long gravitational waves currently observed by the Pulsar Timing Array \cite{NANOGrav:2023hvm}.

However, most of the studies mentioned above, with the exception of \cite{Bolokhov:2023ruj}, focus on perturbations and spectra of massive fields in the background of singular black holes. In \cite{Bolokhov:2023ruj}, the quasinormal modes of a massive scalar field around the regular, ad hoc Bardeen black hole were examined. In this work, we aim to take a further step by investigating the spectrum of a massive scalar field near other regular black holes with an asymptotically Minkowski core, incorporating the effects of an external magnetic field. 

It is worth noting that the literature on perturbations and quasinormal modes of regular black holes is now quite extensive (see, for instance, \cite{Bronnikov:2012ch,Flachi:2012nv,Fernando:2012yw,Toshmatov:2015wga,Toshmatov:2019gxg,Rayimbaev:2022mrk,Konoplya:2023ahd,Konoplya:2023aph} and references therein). However, these studies generally do not consider perturbations of massive fields.

The regular black hole of interest in our study was introduced in \cite{Simpson:2019mud} and differs from the well-known Bardeen and other regular black hole models in several respects.

The paper is organized as follows: Section \ref{Schwarzschild-like} provides an overview of Schwarzschild-like spacetime and discusses the associated magnetic field configuration. In Section \ref{chsf}, we analyze the dynamics of a charged scalar field within this framework. Section \ref{numr} presents the numerical results obtained using both the WKB approximation and the time-domain analysis.

\section{Schwarzschild-like Black Hole \label{Schwarzschild-like}}

The gravitational field of a Schwarzschild-like compact object in Boyer–Lindquist coordinates can be expressed through the following line element~\cite{Simpson:2019mud}:
\begin{align} \label{eq:metric}
ds^2 = -f \, dt^2 + f^{-1} \, dr^2 + r^2 \, (d\theta^2 + \sin^2\theta \, d\phi^2)\ ,   
\end{align}
where the metric function \( f \) is defined as:
\begin{eqnarray}
f = 1 - \frac{2Me^{-\text{a}/r}}{r}\ . \label{eq:delta}
\end{eqnarray}
In this context, the parameter \( M \) represents the mass of the black hole, while a is a deviation parameter introduced by Simpson and Visser \cite{Simpson:2019mud,Simpson:2021dyo}. It is important to note that the spacetime metric described above corresponds to the standard Schwarzschild black hole in general relativity when \( \text{a} \to 0 \).

\subsection{Magnetic Field Configuration}

In realistic astrophysical scenarios, the magnetic field configuration near a gravitationally compact object is highly complex. However, for simplicity, one can consider an analytical expression for the magnetic field. A straightforward approach is provided by Wald~\cite{Wald1974PRD}. According to this approach, the black hole is placed in an asymptotically uniform magnetic field, and the exact analytical expression for the vector potential in Schwarzschild space is given as:
\begin{align}\label{Uniform0}
A_{\phi,Sch} = \frac{1}{2}B r^2\sin^2\theta\ ,
\end{align}
where \( B \) is the magnetic field strength. Note that the expression (\ref{Uniform0}) is independent of the mass of the Schwarzschild black hole and fully satisfies Maxwell's equations in curved spacetime, given as:
\begin{align}\label{ME}
\nabla_\alpha F^{\alpha\beta} = 0\ . 
\end{align}
Similarly, in the background of the Schwarzschild-like spacetime, Maxwell's equation can be analytically solved, and the expression for the vector potential can be found as:
\begin{align}\label{Sol}
A_\phi = \frac{1}{2}B\psi(r)\sin^2\theta\ ,    
\end{align}
which is similar to the solution in the Schwarzschild spacetime in \eqref{Uniform0}, but with a new radial function \(\psi(r)\) substituting \( r^2 \). By inserting equation (\ref{Sol}) into (\ref{ME}), one can obtain:   
\begin{align}\label{ME0}
r^2\left[f\psi'(r)\right]' - 2\psi(r) = 0\,   
\end{align}
where a prime denotes the derivative with respect to the radial coordinate. The solution to equation \eqref{ME0} can be expressed as \(\psi(r) = C(r^2 -2 \text{a} M)\), where \( C \) is a constant of integration, which can be set to 1. Finally, the exact analytical solution of Maxwell's equation for the vector potential near the Schwarzschild-like black hole can be found as:
\begin{align}
A_\phi = \frac{1}{2}B\left(r^2 - 2 \text{a} M\right)\sin^2\theta\ ,
\end{align}
%
%

\begin{figure}
\includegraphics[width=0.92\linewidth]{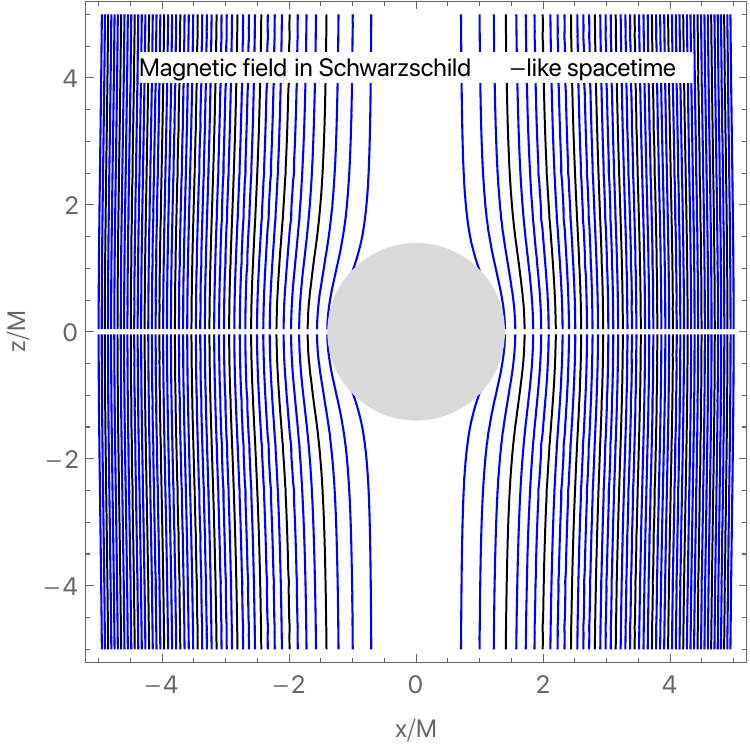}
\caption{The magnetic field lines in the vicinity of the Schwarzschild-like spacetime ($\text{a}/M=0.5$).\label{mag}}
\end{figure}

To study the magnetic field configuration in the vicinity of the black hole, we can analyze the magnetic field lines represented by the equation \( A_\phi = \text{const} \). Figure \ref{mag} provides a visualization of the magnetic field lines near the Schwarzschild-like black hole. It is apparent that due to the additional term in the metric function, the magnetic field lines extend outward from the black hole in a more uniform manner.

\section{charged scalar field}\label{chsf}
The relativistic Klein-Gordon equation for a massive, charged scalar field $\Psi$, in the presence of an electromagnetic field, is expressed as follows:

\begin{align}\label{kge}
    g^{\alpha\beta}(\Delta_\alpha-iqA_\alpha)(\Delta_\beta-iqA_\beta)\Psi-\mu^2\Psi=0\ ,
\end{align}
Here, \( \mu \) represents the mass of the scalar field, \( q \) is the charge coupling constant between the scalar and electromagnetic fields,\( \nabla_{\alpha} \) denotes the covariant derivative and $i$ is imaginary number. Although separating variables in equation (\ref{kge}) is quite challenging, we can simplify the problem by applying the following physically reasonable assumptions:

- Lorentz gauge condition for the vector potential: \( \nabla_{\alpha} A^{\alpha} = 0\);

- In the weak interaction limit, higher-order terms, such as \( q^2 B^2 \), can be neglected, i.e., \( q^2 B^2 \rightarrow 0 \). Then, eq.\eqref{kge} becomes
\begin{align}
  \frac{1}{\sqrt{-g}}\partial_\alpha(\sqrt{-g}g^{\alpha\beta}\partial_\beta\Psi)-2iqA^{\alpha}\partial_{\alpha}\Psi-\mu^2\Psi=0\ ,   
\end{align}

We can write the solution as:
\begin{align}
\Psi(t,r,\theta,\phi) =e^{-i\omega t}Y_{lm}(\theta,\phi)\frac{R(r)}{r} .
\end{align}
With the use the following notation:
\begin{align}
  \nabla^2_{\Omega}= \frac{1}{\sin{\theta}}\partial_{\theta}\left(\sin{\theta}\partial_{\theta}\right)+\frac{1}{\sin^2{\theta}}\partial^2_{\phi} 
\end{align}
one can then obtain the following
\begin{align}
     \nabla^2_{\Omega}\Psi=-l(l+1)\Psi.
\end{align}

Then, in tortoise coordinate system with $dx=dr/f\ ,$ one can write
\begin{align}\label{RVn}
    \left[\frac{d^2}{dx^2}+\omega^2-V\right]R(r)=0
\end{align}
where the effective potential is defined as 
\begin{align}\label{Vn}
    V=f\left[\frac{l(l+1)}{r^2}+\frac{f'}{r}+mqB\frac{2\text{a}M}{r^2}+\mu^2_{eff}\right]
\end{align}
with
\begin{align}
\quad\mu^2_{eff}=\mu^2-mqB\ .
\end{align}
\begin{figure}
    \centering
    \includegraphics[width=0.92\linewidth]{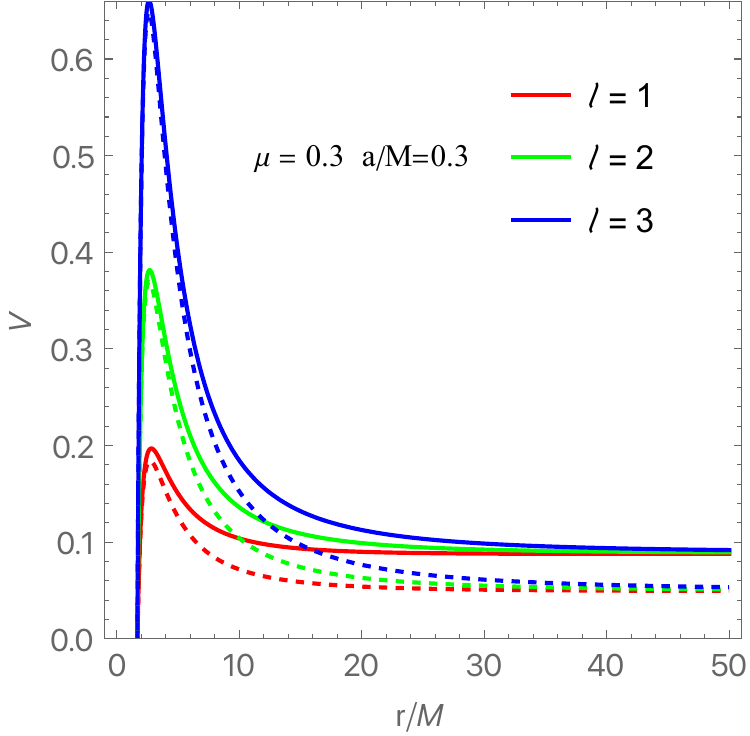}
    \includegraphics[width=0.92\linewidth]{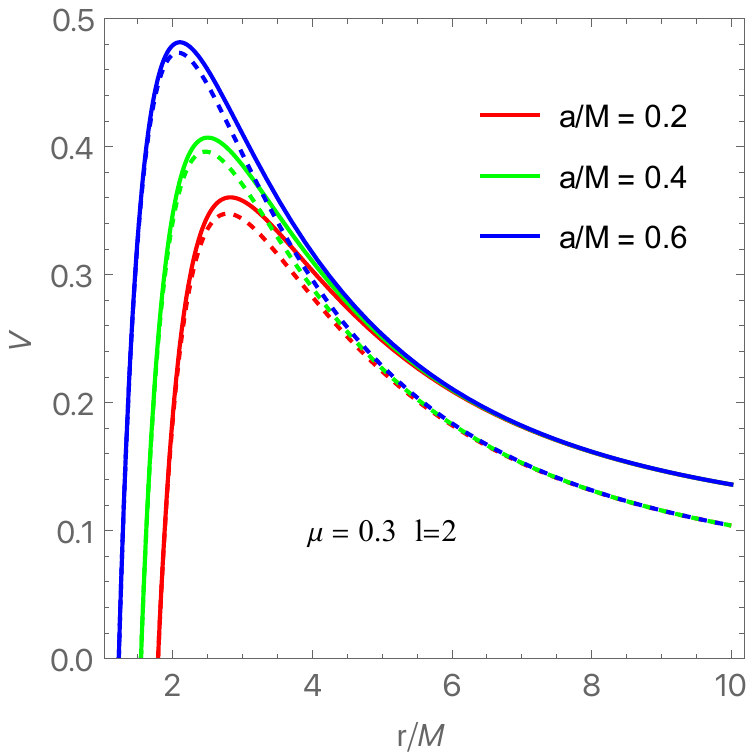}
    \caption{Radial profile of effective potential. Solid  lines for neutral ($qB=0$) particle, dashed lines for charged ($qB=0.2$) particle.}
    \label{nbmVr}
\end{figure}

\begin{table*}[h]
    \centering
    \begin{tabular}{c|c|c|c|c|c|c}\hline\hline
        $l$ & m & a = 0, $\mu=0$, $qB=0$ & a = 0.3, $\mu=0$, $qB=0$ & 
        a = 0.3, $\mu=0.1$, $qB=0$&
        a = 0.5, $\mu=0.1$, $qB=0.1 $&a=0.3, $\mu=0.0$, $qB=0.45 $
        \\
        \hline
         0 & 0 & 0.111946 -0.104579  $i$ &0.125072 - 0.106152$i$& 0.127077 - 0.0992083$i$&0.139158 - 0.0990972$i$& 0.124967-0.105965$i$ \\\hline
         &1& & & &0.364317 - 0.10033$i$&0.264653-0.136211 $i$\\
         1&0&0.292931 - 0.0976602 $i$&0.328766 - 0.100271 $i$&0.332542 - 0.0981747 $i$&0.366747 - 0.0987235$i$&0.328762-0.100294 $i$\\
         &-1 & & & &0.36919 - 0.0970796$i$\\\hline
         &2 & & & &0.598773 - 0.100327$i$&0.452369-0.12924 $i$\\
         &1& & & &0.600472 - 0.0997012$i$&0.496175-0.115227 $i$ \\
         2&0&0.483643 -0.096759 $i$&0.542474 - 0.0994879 $i$&0.54509 - 0.0986551 $i$&0.602177 - 0.0990695$i$&0.542474-0.099488 $i$\\
         &-1& & & &0.603889 - 0.0984322$i$&0.591758-0.081276 $i$\\
         &-2 & & & &0.605608 - 0.0977892$i$&0.644421-0.059118 $i$
         \\\hline
         &3 & & & &0.835555 - 0.100175$i$&0.655526-0.123318 $i$\\
         &2& & & &0.836821 - 0.0998474$i$&0.688443-0.115782 $i$\\
         &1& & & &0.838089 - 0.0995185$i$&0.722359-0.107792 $i$\\
         3&0&0.675366 - 0.0964997 $i$&0.757385 - 0.0992644 $i$&0.75933 - 0.0988275 $i$&0.839361 - 0.0991881$i$&0.757385-0.0992643 $i$\\
         &-1 & & & &0.840635 - 0.0988562$i$&0.793655-0.0900815 $i$\\
         &-2& & & &0.841912 - 0.0985228$i$&0.831329-0.080071 $i$\\
         &-3& & & &0.843192 - 0.0981878$i$&0.870599-0.0689605 $i$\\
         \hline\hline
    \end{tabular}
    \caption{Fundamental quasinormal modes of charged scalar field for Schwarzschild-like black hole obtained by the 6th order WKB ($\tilde{n}=3$, $\tilde{m}=3$, $M=1$).}
    \label{tabwkb}
\end{table*}

\begin{table*}[h]
    \centering
    \begin{tabular}{c|c|c|c|c|c}\hline\hline
        $l$ & m & 6th order WKB($\tilde{n}=3, \tilde{m}=3$) &
        time domain &Relative error(Re($\omega$))& Relative error(Im($\omega$))
        \\
        \hline
         0 & 0 &0.127077 - 0.0992083$i$&&&\\\hline
         &1&0.32919 - 0.100268$i$&&&\\
         1&0&0.332542 - 0.0981747$i$ &$0.332491 - 0.0987287 i$&-0.015\%&-0.56\%\\
         &-1&0.33591 - 0.0960409$i$&&&\\\hline
         &2&0.54038 - 0.100314$i$&&\\
         &1&0.542732 - 0.0994874$i$& &&\\
         2&0&0.54509 - 0.0986551$i$&$0.54509 - 0.0986532 i$&0.00\%&0.0019\%\\
         &-1&0.547456 - 0.0978167$i$&&\\
         &-2&0.549829 - 0.0969722$i$&&\\\hline
         &3&0.754058 - 0.100133$i$ &$0.75407 - 0.100128 i$&0.0016\%&0.0049\%\\
         &2&0.755813 - 0.0996994$i$ & $0.755825 - 0.0996941 i$&0.0016\%&0.0053\%\\
         &1&0.75757 - 0.0992642$i$& $0.757582 - 0.0992592 i$&0.0016\%&0.0050\%\\
         3&0&0.75933 - 0.0988275$i$& $0.759342 - 0.0988225 i$&0.0016\%&0.0051\%\\
         &-1&0.761093 - 0.0983891$i$& $0.761106 - 0.0983841 i$&0.0017\%&0.0051\%\\
         &-2&0.76286 - 0.0979492$i$& $0.762872 - 0.0979442 i$&0.0016\%&0.0051\%\\
         &-3&0.764629 - 0.0975076$i$&$0.764642 - 0.0975025 i$&0.0017\%&0.0052\%\\
         \hline\hline
    \end{tabular}
    \caption{Fundamental quasinormal modes of charged scalar field for Schwarzschild-like black hole obtained by the 6th order WKB (a= 0.3, $\mu=0.1$, $qB = 0.1$, $M=1$).}
    \label{tabwkb1}
\end{table*}

\section{Numerical Results}\label{numr}

 In this section, we briefly present the results of numerical calculations of the quasinormal modes for a Schwarzschild-like black hole immersed in an external asymptotically uniform magnetic field. To solve the Schrödinger-like wave equation (\ref{RVn}), we impose the following boundary conditions: the wave exhibits purely incoming behavior at the event horizon and purely outgoing behavior at spatial infinity, as
\[
R(r) = 
\begin{cases} 
e^{-i\omega x}, & \text{as} \, x \to -\infty \, (r \to -\frac{\text{a}}{W\left(-\frac{\text{a}}{2}\right)}), \\
e^{i\chi x}, & \text{as} \, x \to \infty \, (r \to \infty),
\end{cases}
\]
where \( \chi = \sqrt{\omega^2 - \mu_{eff}^2} \).
\subsection{WKB method}

For frequency-domain analysis, we employ the semi-analytical WKB method \cite{1985ApJ...291L..33S,Froeman:1992gp,PhysRevD.96.024011,Konoplya:2019hlu,Konoplya:2003ii}. This method involves expanding the solution at both infinities in a WKB series and matching these asymptotic expansions with a Taylor series near the peak of the effective potential. Additionally, as suggested in \cite{PhysRevD.96.024011}, we enhance the WKB expansion by representing it with the Padé approximation, which significantly improves the accuracy of the WKB method in most cases.

The higher order WKB formula is given by \cite{Konoplya:2019hlu}:

\begin{align}\label{wkbo}
&\omega^2 = V_0 + A_2(K^2) + A_4(K^2) + A_6(K^2)+\dots\nonumber\\&- iK \left( -2V_1 + A_3(K^2) + A_5(K^2) + A_7(K^2) + \dots \right)
\end{align}

where $ K = n + 1/2$ , with $n = 0, 1, 2, 3, \dots$. 

Corrections $A_k(K^2)$ of order \( k \) to the eikonal formula are polynomials in \( K^2 \) with rational coefficients, depending on the values \( V_2, V_3, \dots \) of higher derivatives of the potential \( V(r) \) at its maximum. To further improve the accuracy of the WKB formula, we use the procedure proposed by Matyjasek and Opala \cite{PhysRevD.96.024011}, which involves applying the Padé approximation. For a given order \( k \) of the WKB formula (\ref{wkbo}), we define a polynomial \( P_k(\epsilon) \) as follows:
\begin{align}
    &P_k(\epsilon) = V_0 + A_2(K^2) \epsilon^2 + A_4(K^2) \epsilon^4 + A_6(K^2) \epsilon^6 + \dots \nonumber\\&- iK \left( -2V_1 \epsilon + A_3(K^2) \epsilon^3 + A_5(K^2) \epsilon^5 + \dots \right)    
\end{align}
and the squared frequency is then obtained by setting \( \epsilon = 1 \):

\begin{align}
    \omega^2 = P_k(1).
\end{align}

To improve the approximation of \( P_k(\epsilon) \), we apply the Padé approximation:

\begin{align}
    \tilde{P}_{\tilde{n}/\tilde{m}}(\epsilon) = \frac{Q_0 + Q_1 \epsilon + \dots + Q_{\tilde{n}} \epsilon^{\tilde{n}}}{R_0 + R_1 \epsilon + \dots + R_{\tilde{m}} \epsilon^{\tilde{m}}},    
\end{align}

where \( \tilde{n} + \tilde{m} = k \), ensuring that near \( \epsilon = 0 \), \( \tilde{P}_{\tilde{n}/\tilde{m}}(\epsilon) - P_k(\epsilon) = O(\epsilon^{k+1}) \).

For finding the fundamental mode (\( n = 0 \)), Padé approximation with \( \tilde{n} \approx \tilde{m} \) generally yield the best results. In \cite{PhysRevD.96.024011}, the approximation \( P_{6/6}(1) \) and \( P_{6/7}(1) \) were compared for this purpose. The 6th-order WKB formula \( P_{6/0}(1) \) is commonly used, but as noted in \cite{Konoplya:2019hlu}, even the Padé approximation \( P_{3/3}(1) \) of the 6th order often yields a more accurate value for the squared frequency than \( P_{6/0}(1) \). In this work, we will use the 6th WKB expansions.

Using the sixth-order WKB approximation, we compute the quasinormal mode frequencies of electrically charged scalar perturbations in the field of a magnetized Schwarzschild-like black hole. Table \ref{tabwkb} displays the real and imaginary components of these frequencies for various azimuthal (\(m\)) and multipole (\(l\)) numbers, as well as for different values of the parameter \(qB\) and the spacetime parameter \(a\). Furthermore, in Table \ref{tabwkb1}, we compare the results obtained using the sixth-order WKB method with those derived from the time-domain method. The relative errors in the real (third column) and imaginary (fourth column) parts are shown to be negligible, demonstrating the high accuracy of the WKB results.

\begin{figure*}
    \centering
    \includegraphics[width=0.45\linewidth]{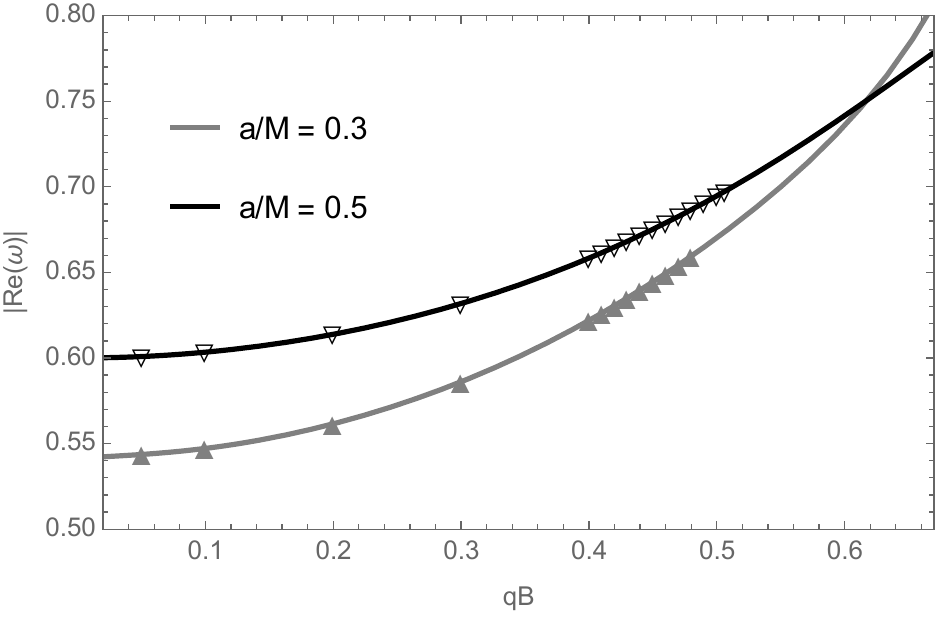}
    \includegraphics[width=0.45\linewidth]{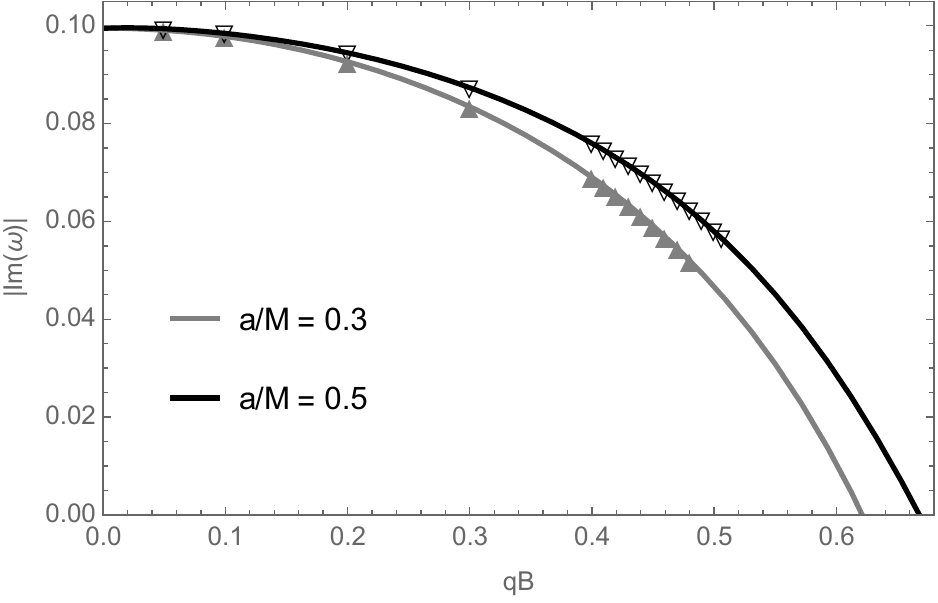}
    \caption{Best fitting polynomial functions of WKB data for different values of spacetime values($l=2,m=-2$).}
    \label{imreo}
\end{figure*}
Fig.\ref{imreo} illustrates the best-fit polynomial function for data obtained using the WKB method, applied to a massless scalar field in an external magnetic field. The figure shows two curves, gray and black, corresponding to different spacetime parameter values (\(a/M = 0.3\) and \(a/M = 0.5\), respectively). The imaginary parts of the quasinormal mode frequency \(\omega\) for these parameter values are shown on the right-hand side. Each curve intersects the zero value of \(qB\) at approximately 0.611711 (gray line) and 0.651842 (black line), respectively.  

\begin{figure*}
    \centering
    \includegraphics[width=0.45\linewidth]{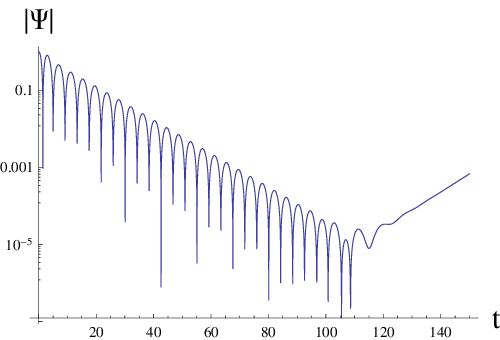}
    \includegraphics[width=0.45\linewidth]{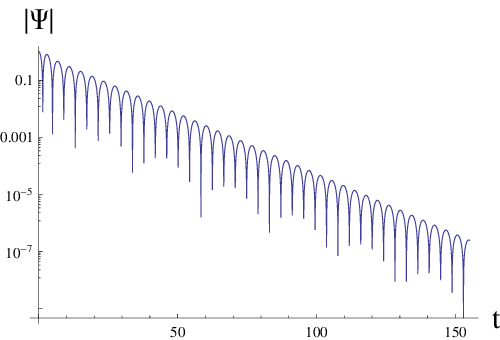}
    \caption{Semi-logarithmic plots of time-domain profiles for $m=3$ (left) and $m=-3$ (right) perturbations; $l=3$, $q B =0.1$, a = 0.3, $\mu =0.1$, $M=1$. The instability at late times for $m>0$ is the artifact of the approximation for the effective potential which is valid only until some distance from the black hole (see discussion in \cite{Kokkotas:2010zd}). At the ringdown phase the WKB data is reproduced with high accuracy. }
    \label{TD1}
\end{figure*}

\begin{figure*}
    \centering
    \includegraphics[width=0.45\linewidth]{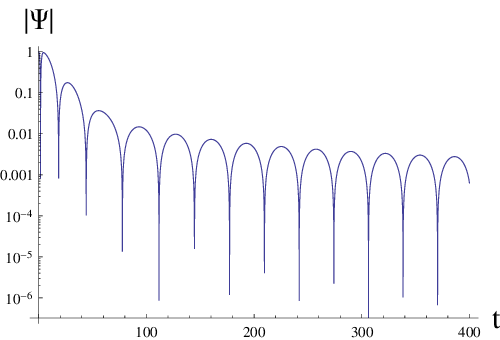}
    \includegraphics[width=0.45\linewidth]{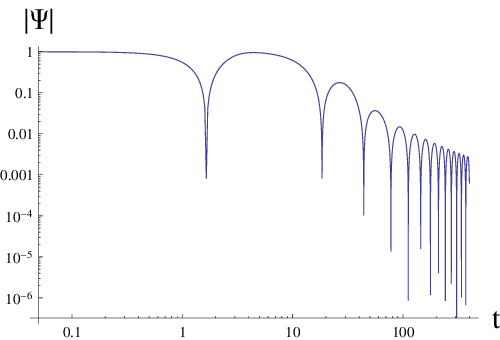}
    \caption{Semi-logarithmic plots of time-domain profiles (left) and logarithmic plot (right); $l=m=0$, $q B =0.1$, a = 0.3, $\mu =0.1$, $M=1$. The short period of quasinormal ringing is quickly changed by an asymptotic tails, making it unable to extract the frequency from the profile with sufficient accuracy.}
    \label{TD2}
\end{figure*}

\begin{figure*}
    \centering
    \includegraphics[width=0.45\linewidth]{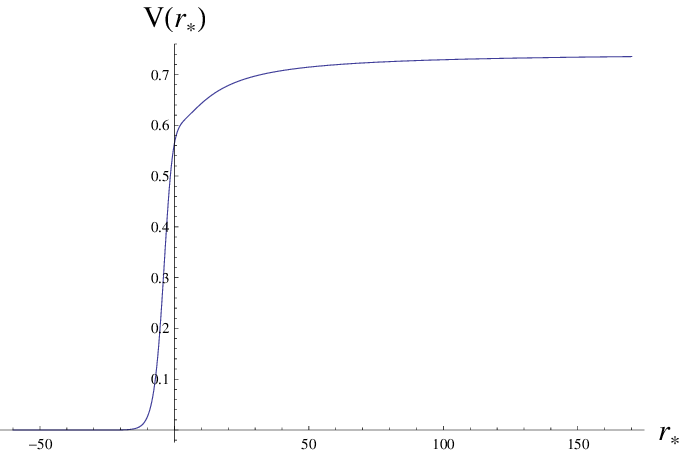}
    \includegraphics[width=0.45\linewidth]{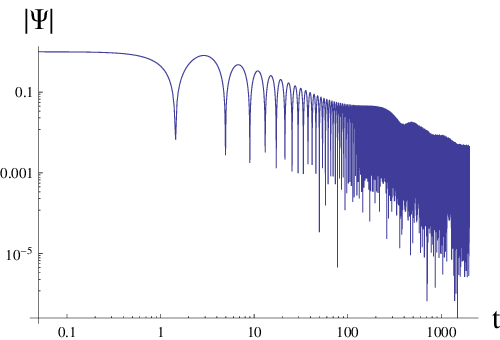}
    \caption{Effective potential and logarithmic plots of time-domain profiles for $m=-2$  perturbations; $l=2$, $q B =0.61$, a = 0.3, $\mu =0$, $M=1$.}
    \label{TD3}
\end{figure*}

The WKB data is checked here by time-domain integration. However, two aspects must be taken into account for such a comparison. First of all, the period of quasinormal ringing is very short for $\ell=0$ perturbations because it is quickly followed by oscillatory power-law tails, as shown in Fig.~\ref{TD2}. Therefore, it is difficult to extract the frequency from the time-domain profile with sufficient accuracy in this case. The second aspect is the growth of the perturbation at late times Fig. \ref{TD1}. One could think that this means an instability caused by the negative value of the effective potential at large $r$ when $m>0$. However, as shown in \cite{Kokkotas:2010zd}, the effective potential itself is valid only up to some range that is inversely proportional to the asymptotic value of the magnetic field $B$. Thus, we conclude, in the same way as in \cite{Kokkotas:2010zd}, that the instability is false and is simply an artifact of treating the effective potential which is determined only in some range near the black hole valid throughout the whole space instead. Even in that case, the quasinormal ringing period before the ``instability'' is in very good agreement with the WKB data.

\subsection{The time-domain integration}

We rewrite the wave equation (\ref{RVn}) without assuming the stationary ansatz ($\Psi$ $\sim$ $e^{-i\omega t}$) as:

\begin{align}\label{nwe}
\frac{\partial^2 \Psi}{\partial t^2} - \frac{\partial^2 \Psi}{\partial x^2} + V(t, x) \Psi = 0   
\end{align}
where \(x\) is the tortoise coordinate. The method for integrating this wave equation in the time domain was developed by Gundlach, Price, and Pullin \cite{Gundlach:1993tp}. We now express equation (\ref{RVn}) in terms of the light-cone coordinates \( du = dt - dx \) and \( dv = dt + dx \), resulting in:

\begin{align}
   \left(4 \frac{\partial^2}{\partial u \partial v} + V(u, v)\right) \Psi(u, v) = 0
\end{align}

This equation is solved numerically using a discretization scheme as follows \cite{Gundlach:1993tp}:
\begin{eqnarray}
\Psi\left(N\right)&=&\Psi\left(W\right)+\Psi\left(E\right)-\Psi\left(S\right)\nonumber\\
&&- \Delta^2V\left(S\right)\frac{\Psi\left(W\right)+\Psi\left(E\right)}{4}+{\cal O}\left(\Delta^4\right),\label{Discretization}
\end{eqnarray}
Here, we have the following points for the integration scheme: $N\equiv\left(u+\Delta,v+\Delta\right)$, $W\equiv\left(u+\Delta,v\right)$, $E\equiv\left(u,v+\Delta\right)$,  $S\equiv\left(u,v\right)$,  and  \( \Delta \) is a constant representing the separation between neighboring grid points (further details can be found in references \cite{Gundlach:1993tp,PhysRevD.97.084058}). Initial conditions are specified on the two null surfaces, \( u = u_0 \) and \( v = v_0 \). For computations, we assume that the initial perturbation is a Gaussian pulse centered around \( v_c \) with width \( \sigma \), given by:

\begin{align}
    \Psi(u = u_0, v) = A \exp \left( -\frac{(v - v_c)^2}{\sigma^2} \right).
\end{align}
\textbf{}

Then, utilizing the Prony method \cite{Prony1795} to represent the signal as a sum of exponential terms with associated weights , we extract the dominant quasinormal frequencies from the time-domain profile. The time-domain integration approach, known for its high precision, has been extensively employed in numerous studies (see,  for example, \cite{Lin:2024ubg,Bolokhov:2023dxq,Cuyubamba:2016cug,Dubinsky:2024gwo,Malik:2024tuf,Skvortsova:2024atk,Al-Badawi:2024kdw,Aneesh:2018hlp}. Consequently, we will not elaborate on it further in this work.

As shown in Fig.~\ref{TD3}, when $\ell$ and $m$  are non-vanishing, and some values of other parameters are fixed, the asymptotic behavior qualitatively deviates from the usual massive tails, which are oscillatory with a power-law envelope \cite{Hod:1998ra,Koyama:2001qw,Koyama:2001ee,Moderski:2001tk,Gibbons:2008gg,Jing:2004zb,Konoplya:2006gq}. 
In our case, the presence of a magnetic field results in an oscillatory envelope instead of a power-law one, which cannot be easily fitted to a simple analytical formula. A similar behavior was recently observed in the context of environmental effects on the late-time decay of massive fields \cite{Konoplya:2024wds}. In that case, however, the configuration did not involve a magnetic field; the environmental effects were instead modeled by Gaussian bumps in the effective potential. In our case this unusual behavior should not be determined by the limited approximation of the range of the effective potential. We believe this phenomenon occurs due to the specific behavior of the effective potential near the event horizon, as illustrated in Fig.~\ref{TD3}. In terms of the tortoise coordinate, there is a distinct kink near the event horizon. As demonstrated in \cite{Konoplya:2024wds}, deformations in the effective potential close to the horizon can significantly modify the late-time tails of massive fields. Additionally, given that massive tails decay slowly and are expected to contribute to the very long-wavelength radiation observed via the Pulsar Timing Array \cite{Konoplya:2023fmh}, the unusual late-time decay induced by magnetic fields opens up a new avenue for further investigation.

\section{Conclusions}

While the quasinormal modes of regular black holes have been extensively studied in numerous works, no comprehensive investigations have been conducted for regular black holes in the presence of an external magnetic field. In this paper, we address this gap and demonstrate that the magnetic field significantly alters the spectrum of a charged scalar field, leading to the emergence of arbitrarily long-lived quasinormal modes, known as quasi-resonances \cite{Ohashi:2004wr}.

Another distinctive feature of the evolution of perturbations in the presence of a magnetic field is the unusual behavior of asymptotic tails. For certain parameter values, these tails do not exhibit a power-law envelope but instead display an oscillatory envelope. This behavior is most likely due to the peculiar deformation of the effective potential near the event horizon and deserves a detailed investigation in future studies.

This work could be further extended by considering other regular black hole solutions and different magnetic field structures.

\section*{Acknowledgement}
We would like to express our sincere gratitude to Dr. Roman Konoplya for his valuable insights and constructive feedback, which greatly contributed to improving the quality of this work.

\bibliographystyle{apsrev4-2}  
%

\end{document}